\documentclass[conference]{IEEEtran}
\IEEEoverridecommandlockouts
\usepackage{cite}
\usepackage{amsmath,amssymb,amsfonts}
\usepackage{algorithmic}
\usepackage{graphicx}
\usepackage{textcomp}
\usepackage{xcolor}
\usepackage{multirow,tabularx}
\usepackage{soul}
\usepackage{multirow}
\usepackage{xcolor}
\usepackage{subcaption}

\def\BibTeX{{\rm B\kern-.05em{\sc i\kern-.025em b}\kern-.08em
    T\kern-.1667em\lower.7ex\hbox{E}\kern-.125emX}}

\title{ESAI: Efficient Split Artificial Intelligence via Early Exiting Using Neural Architecture Search}

\author{\IEEEauthorblockN{Behnam Zeinali\IEEEauthorrefmark{1},
        Di Zhuang\IEEEauthorrefmark{2}, 
        and~J. Morris Chang\IEEEauthorrefmark{3}}
\IEEEauthorblockA{Department of Electrical Engineering \\
University of South Florida \\ Tampa, Florida 33620\\
Email: \IEEEauthorrefmark{1}behnamz@usf.edu, 
\IEEEauthorrefmark{2}dizhuang@usf.edu,
\IEEEauthorrefmark{3}chang5@usf.edu}}

\begin{document}

\maketitle

\begin{abstract}
Recently, deep neural networks have been outperforming conventional machine learning algorithms in many computer vision-related tasks. However, it is not computationally acceptable to implement these models on mobile and IoT devices and the majority of devices are harnessing the cloud computing methodology in which outstanding deep learning models are responsible for analyzing the data on the server. 
This can bring the communication cost for the devices and make the whole system useless in those times where the communication is not available. 
In this paper, a new framework for deploying on IoT devices has been proposed which can take advantage of both the cloud and the on-device models by extracting the meta-information from each sample's classification result and evaluating the classification's performance for the necessity of sending the sample to the server. Experimental results show that only 40 percent of the test data should be sent to the server using this technique and the overall accuracy of the framework is 92 percent which improves the accuracy of both client and server models. 

{Impact Statement---}In this article, a novel method of actionable intelligence implementation on local mobile devices has been proposed. The proposed framework intelligently decides and controls if the sample should be operated on the local or the server model using the meta-information. The performance of the embedded AI unit is improved by proposing the novel method of neural architecture search technique using the knowledge distillation idea. Also, the efficiency and flexibility of the embedded AI unit have been enhanced by introducing the earlier exit on the client model. Experimental results show the efficiency and effectiveness of the proposed framework. The framework is also implemented on real devices.

\end{abstract}


\section{Introduction}
With the advancement of deep learning, deep neural networks have been utilized in many real-world applications, such as botnet detection, \cite{community}
community detection \cite{botnet_1,botnet_2,community}, active authentication \cite{active_authentication} as well as facial recognition systems \cite{face_1,face_2}. Information technology such as Cloud computing, wireless communication, and the Internet of Things have also progressed and as a result, mobile and IoT devices are ubiquitous these days. The progress that has been occurred in different fields of mobile technologies recently has made mobile devices more practical and desirable. An important one of such technologies is enabling deep learning models running on mobile and IoT devices.

However, deploying Deep Learning applications on mobile and IoT devices is not an easy task to do so. The majority of practical deep learning models are heavy and power-consuming and consequently, it is not possible to implement them on mobile devices. On the other hand, using a lighter model with less power consumption can reduce the accuracy of the model drastically. Considering the trade-off between resource usage efficiency and prediction accuracy is vital for such an application. 
Cloud computing technology can be operated to tackle this problem. For example, Google Cloud AutoML \cite{google} has provided Machine Learning as a Service (MLaS) system that is providing powerful servers for machine learning applications. Using IoT devices, data should be sent to these servers and after performing machine learning tasks, results will be shown to the users. Even though this approach provides better accuracy and less computational energy consumption for mobile devices, it brings the communication cost that needs to be considered. It also requires permanent and appropriate communication resources. Nonetheless, there are some circumstances that communication resources are limited and even not available. For example in some remote or rural areas satellite communication may not be available. 

Consequently, it is very important to design a framework to enable such applications that take advantage of both the client and the server-side. The framework should be a client-server system where the client is a mobile device and the server is cloud computing. For each input sample, The framework intelligently makes a decision to classify it either using the client or server model. By doing so, it will be able to receive the best advantages from the server-provided model and the client model in a very efficient manner with as maximum accuracy as possible. 

There are some challenges in designing such a system. An accurate, compact, and efficient DNN model needs to be deployed on the client-side to have an acceptable accuracy with the least power consumption in any scenarios that satellite network is not available. During the network availability, the framework has to be intelligent enough to choose suitable samples and send them to the cloud network to get better accuracy with lower communication costs. Moreover, since saving energy is very important in mobile devices,  the framework should exit the sample in the earlier stage when the same result can be achieved with fewer layers which can reduce the computational energy consumption. 

The model on the client needs to be compressed enough to be deployable on mobile devices. It should also have appropriate accuracy in such a scenario that satellite resources are not available. A few pieces of research have been done so far to provide compressed and accurate models. Knowledge distillation techniques  \cite{hinton-distilling, ba2014deep, polino2018model}consider a heavy and complex model with acceptable accuracy as a teacher and try to teach or convey the information from this model to a simpler model in terms of layers and size. The simple model which known as the student model has the lass size but about the same accuracy. In this technique, the framework is not able to take advantage of the model on the server which can be a model with very high accuracy without any computational limitation. Split-DNN architecture\cite{jeong2018computation, kang2017neurosurgeon, lane2016deepx, osia2020hybrid} is also used to have a lighter model on the client. In this technique, a complex DNN model is divided into two sections so that the lighter one is used on the client and the heavier one is used on the server. However, these models are highly dependant on the availability of satellite resources and are not able to work when communication is impeded.

In this paper, we propose an intelligent and efficient AI framework using multi-exit DNN. 
This framework is based on a client-server system and intelligently makes a decision about each input data to whether it needs to be sent to the server or can be classified using the client model. This decision-making is based on the amount of the required precision and the energy state of the device and the trade between the two of them needs to be optimized by the framework. About the data that does not need to be sent to the server, the framework decides to exit it from the first exit if it is possible to use less computational energy. 
In order to achieve a light model with acceptable accuracy for the client model, we have used NAS morphism \cite{nas-morphism} algorithm to find an appropriate network. 
By applying the knowledge distillation technique \cite{hinton-distilling}, a new searching strategy has been proposed to get a light model with acceptable accuracy. Then, the obtained model has been converted to a multi-exit model. 
A decision unit for each exit has been proposed separately using meta-information which is responsible to make a decision about each data sample in each exit for sending it to the next stage or the server or showing the classification result to the user.

To evaluate the proposed method, in the experimental results section, first, we evaluate the efficacy of applying the knowledge distillation technique\cite{hinton-distilling} to the morphism-based neural architecture search algorithm \cite{nas-morphism}. 
Then we show that this model provides better meta information for training the decision unit comparing to some of the state-of-the-art models which have been introduced so far for implementing on IoT devices. 
In the end, we evaluate the accuracy and efficiency of the multi-exit framework in saving computational power.

Skin lesion 2019 data set \cite{ISIC_1,ISIC_2,ISIC_3} has been used to evaluate our proposed method. The data set contains 25331 images from eight diseases. The data set has been divided into train, validation, and test set which contains 80, 10, and 10 percent of the data set.

To summarize, the proposed method contributions are:
\begin{itemize}
  \item Designing a novel and intelligent Split-AI architecture to get efficient and accurate deep learning services from the mobile/IoT based applications from both client and server models
  \item Enhancing the design of morphism-based NAS by introducing Knowledge Distillation techniques into its searching strategy to get a mobile-deployable light model
  \item Introducing a multi-exit DNN architecture to enable flexible and efficient tuning trade-off between the resource usage efficiency and prediction accuracy
\end{itemize}

In the section ~\ref{sec:Preliminaries}, we will talk about some preliminary knowledge about NAS algorithms and knowledge distillation techniques. Then In the section ~\ref{sec:methodology}, we introduce the proposed method in more detail step by step. The experimental result will be shown in the section ~\ref{sec:result} . In the section ~\ref{sec:related}, we will take a look at some efforts the have been done so far by researchers to implement deep neural networks on mobile devices.

\begin{figure*}
  \centering
    \includegraphics[width=\linewidth]{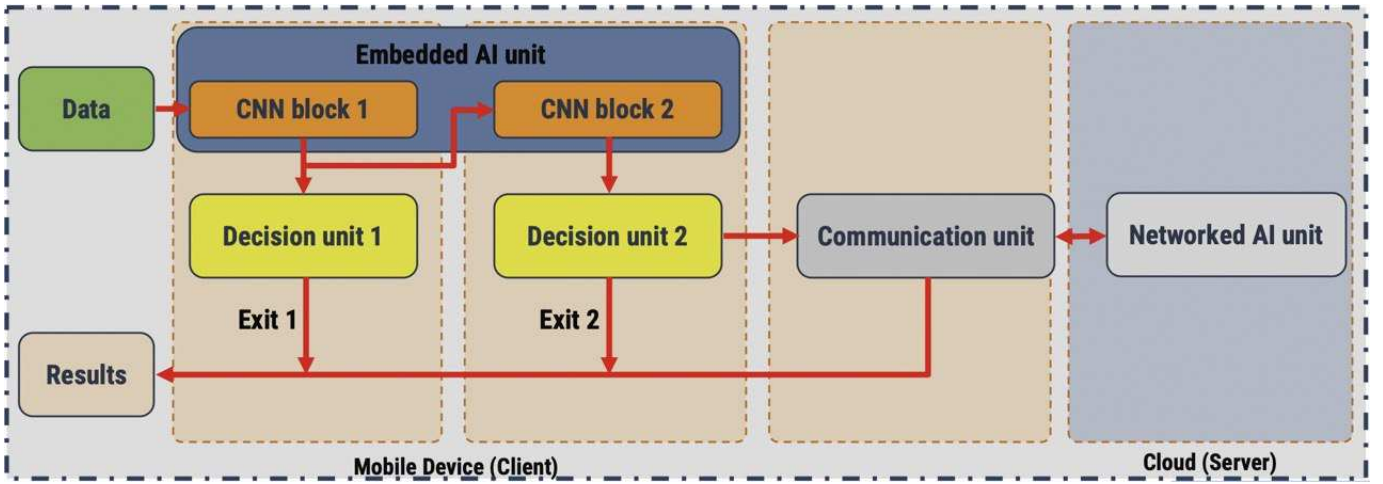}
  \caption{The proposed framework}
  \label{fig:proposed_fw}
\end{figure*}
\section{Preliminaries}\label{sec:Preliminaries}
In this section, we introduce a basic understanding of neural architecture search (NAS) and knowledge distillation techniques (KD).
\subsection{Neural Architecture Search and morphism-based NAS}
Human-made neural network architectures have been progressed tremendously since the Alexnet network was introduced in 2012 \cite{alexNet}. Variety of neural networks with different architectures have been designed by researchers such as VGG19\cite{VGG19}, InceptionV3 \cite{InceptionV3}, Xception\cite{Xception}, DenseNet\cite{DenseNet},  InceptionResNetV2\cite{inceptionv4}, and ResNet-152\cite{Res_Net} while each of them has different combination of convectional layers, fully connected layers, and concatenation and skip layers. These models are general models that have been designed by some experts that have years of experience in the field of machine learning and feature engineering. Even though these kinds of designing are time-consuming and tedious to do so, it requires knowledge to design a different architecture for different data set and a target.

To address this problem, a new technique called Neural Architecture Search\cite{NasNet} has been introduced to overcome these mentioned drawbacks. It tries to search and find a good architecture for a specific target in an automatic manner which can lead to finding a model with better results comparing to man-made models in the classification tasks. For doing so, two important definitions should be clarified for running such an algorithm. First, what is the search space, where and how the algorithm is going to connect different architecture together reasonably, and second, how it will be able to evaluate each new architecture and find the optimum one with the best results? Different techniques for addressing these two strategies have been introduced recently. Liu et al\cite{Pnasnet} have introduced cell-based NAS algorithms in which the number of different architecture cells is connected together to define a search space in every try and an evolutionary algorithm has been used to evaluate different architectures.  Liu et al \cite{dart}
used the same strategy for search space but defined a bilevel optimization system to evaluate each architecture.

Neural Architecture Search method based on the morphism algorithm has been introduced by Elsken et al\cite{nas-morphism}. In their work, for defining the search space, they have introduced a morphism based method in which at first, the algorithm defines an initial architecture, then it tries to extend this architecture by randomly adding different layers such as fully connected, convolution, skip, and concatenation layers to its layers. The layers are being selected in a random fashion. For addressing the second definition, the algorithm uses a hill-climbing method to evaluate each new architecture. hill climbing is a greedy algorithm that compares each new model to the previous one and replaces it if the new one shows better performance.
\subsection{Knowledge Distillation}
The majority of man-made models, as well as models that have been obtained by NAS algorithms, are cumbersome which makes them efficiently unreasonable to be implemented on IoT devices due to their power and storage consumption. Most of the energy in IoT devices is consumed by memory references and reducing the memory size of the model can lead to less storage and consequently less power consumption. Different techniques of model compression and acceleration have been proposed to overcome these problems. Knowledge distillation\cite{hinton-distilling} and model compression\cite{compression} techniques can be used to reduce the model size while maintaining the same accuracy. Attentions to knowledge distillation have been increased rapidly in recent years due to its simplicity and effectiveness\cite{kd-survey}. This technique trains a single small model by using transferred knowledge from a cumbersome model and a model with less model size but approximately the same accuracy can be obtained. In order to do so, instead of training the smaller model using the hard labels of the data set, it is being trained using the softened labels of each data sample using the output of Softmax calculation of that sample from the cumbersome model. In this way, more information has been transferred to the smaller model without transferring the noise and unnecessary data. The amount of data that have been transferred to the smaller model can be controlled by the temperature value (T) in the following equation:

\begin{equation}
S_i = \frac{\exp{({z_i}/{T}})}{\sum_{j}{\exp{({z_j}/{T}})}}  
\end{equation}
A bigger value for T means that more information is being transferred to the smaller model. The smaller model does not need to worry about overfitting since the cumbersome model has already dealt with that problem.

\section{Methodology}\label{sec:methodology}

\subsection{Overview}
For a client-server Split-AI Architecture that wants to take advantage of both the model on the client (mobile and IoT devices) and the model on the server(a complex model with high accuracy), it is of vital importance to make an accurate decision about whether a sample should be sent to the server or be classified using the client model. For a client model with 80 percent accuracy, in the ideal scenario, 20 percent of samples (falsely classified samples) should be determined and sent to the server. The model on the server has no limitation in terms of storage and computational cost and can be a model with better performance. By sending data in an intelligent manner, not only the framework can take advantage of the client model as much as possible, but also it can use the communication cost only for those samples that need to be classified by a better model on the server. In this manner, overall accuracy using both client and server models can be improved while using fewer communication costs. The decision unit tries to find falsely classified samples after classification using the meta-information. It learns how to make a decision about test samples by being trained using the labeled meta information that has been obtained from applying the validation data set to the client model. To reduce the computational power, another exit can be extracted from the client model and use a similar idea to prevent data to get through the whole layers of the model. Therefore, not only the model on the client side needs to be small to have less power consumption, but also it has to be accurate enough even in the earlier exit to produce appropriate information for training the decision unit. To get an appropriate client model, our proposed method improves the NAS morphism algorithm in terms of accuracy by applying the knowledge distillation technique to its searching strategy. In section~\ref{sec:Split-AI}, we introduce our Split-AI framework in more detail. Then in the section~\ref{sec:Client Model}, the proposed method for getting the client model will be presented. In section~\ref{sec:Multi-exit}, a multi-exit idea will be explained.

\subsection{Split-AI framework}\label{sec:Split-AI}
The proposed Split-AI framework has three different parts. A client model is a light model with acceptable accuracy. A decision unit that decides about each sample after being classified and a communication unit that is responsible for sending the data to the server, getting the results, and show them to the user. Figure~\ref{fig:proposed_fw} shows the proposed framework. The Decision unit plays an important role in the framework. It is responsible to identify samples that have been classified incorrectly by the client model and send them to the communication section for being classified by the server. It is a light machine learning model that is trained by using meta-information.

To do so, after classifying some samples with the machine learning model, meta-information has been extracted from the result and since we have the label of that sample, the Decision Unit is being trained about what decision it should make about that sample. For testing, those samples that have more uncertainty are separated and sent to the server. Certainty means that it is highly probable that the machine learning model has classified those samples correctly. Based on the uncertainty, a parameter name decision unit's sensitivity is defined which is a number between 0 and 1. In our work, the server model plays an expert role. Indeed, the decision unit will show the classification result to the user only when the certainty of its decision exceeds its value of sensitivity, otherwise, it will send the sample to the communication stage to be sent to the server.

A similar idea has been used in active learning\cite{activelaerning}. Labeling the data by an expert is one of the most important and tedious work in the field of machine learning. When you have a new data set with unlabeled data, you need to label them based on the class that each sample belongs to. In the case of the medical data set, for labeling data, hiring an expert is vital which makes it expensive and important. When the data set is really large, it is not logically and financially acceptable to send the whole samples to the experts to be labeled. One of the ideas here is determining and sending some samples to the expert which can lead to better results. Here, samples with higher uncertainty are determined and sent to the expert using a similar idea.

\subsubsection{Meta Information}
Different types of meta-information can be extracted from the machine learning model output to identify the uncertainty of the results. In our work, we have extracted maximum probability, least confidence, entropy, and standard deviation as meta information. The equations for all of them can be defined as below respectively:

\begin{equation}\label{second_eq}
MP = max(P_i)
\end{equation}
\begin{equation}\label{third_eq}
LC = (P_s(0) - P_s(1)) 
\end{equation}
\begin{equation}\label{forth_eq}
Entropy = \sum_{i}{P_i* \log{P_i} }
\end{equation}
\begin{equation}\label{fifth_eq}
std = \sqrt {\frac{\sum{(P_i - \mu)^2}}{N}}
\end{equation}

where $P_s = sort(P_i)$ , $P_i$ is the probability of the client output for each class $i$, $\mu$ is the mean value of $P_i$ and $N$ is the number of classes. Imagine a classifier with three classes and sampleA and sampleB have been applied to this classifier and we want to decide which one of them has more uncertainty and can be a better candidate to be sent to the server instead of on-device classification. Table ~\ref{table:metainformation} shows the extracted meta information from two example samples. For example, the least confidence of sampleA after being classified is $0.9-0.09 = 0.81$ while sampleB is $0.5 - 0.3 = 0.2$ which means sampleB has higher uncertainty and it is more prone to be sent to the server. By computing other values (entropy, maximum probability, and standard deviation which are the last three rows of the table ~\ref{table:metainformation}), sampleB is an appropriate sample to be sent to the server.

\subsubsection{Decision Unit}
\begin{figure*}[h!]

  \centering
  \begin{subfigure}[b]{0.45\linewidth}
    \includegraphics[width=\linewidth]{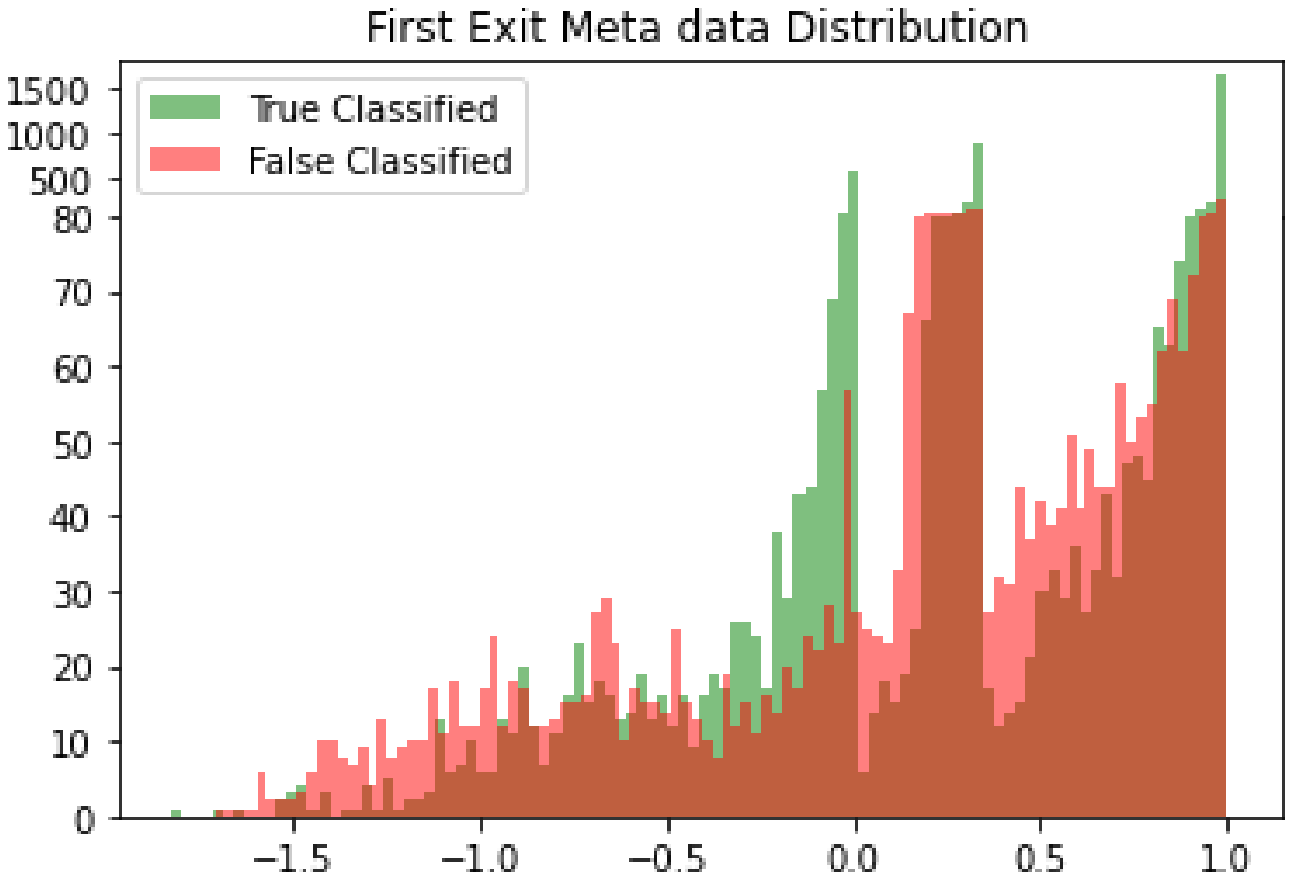}
     \caption{First Exit Meta Data Distribution}
  \end{subfigure}
  \begin{subfigure}[b]{0.45\linewidth}
    \includegraphics[width=\linewidth]{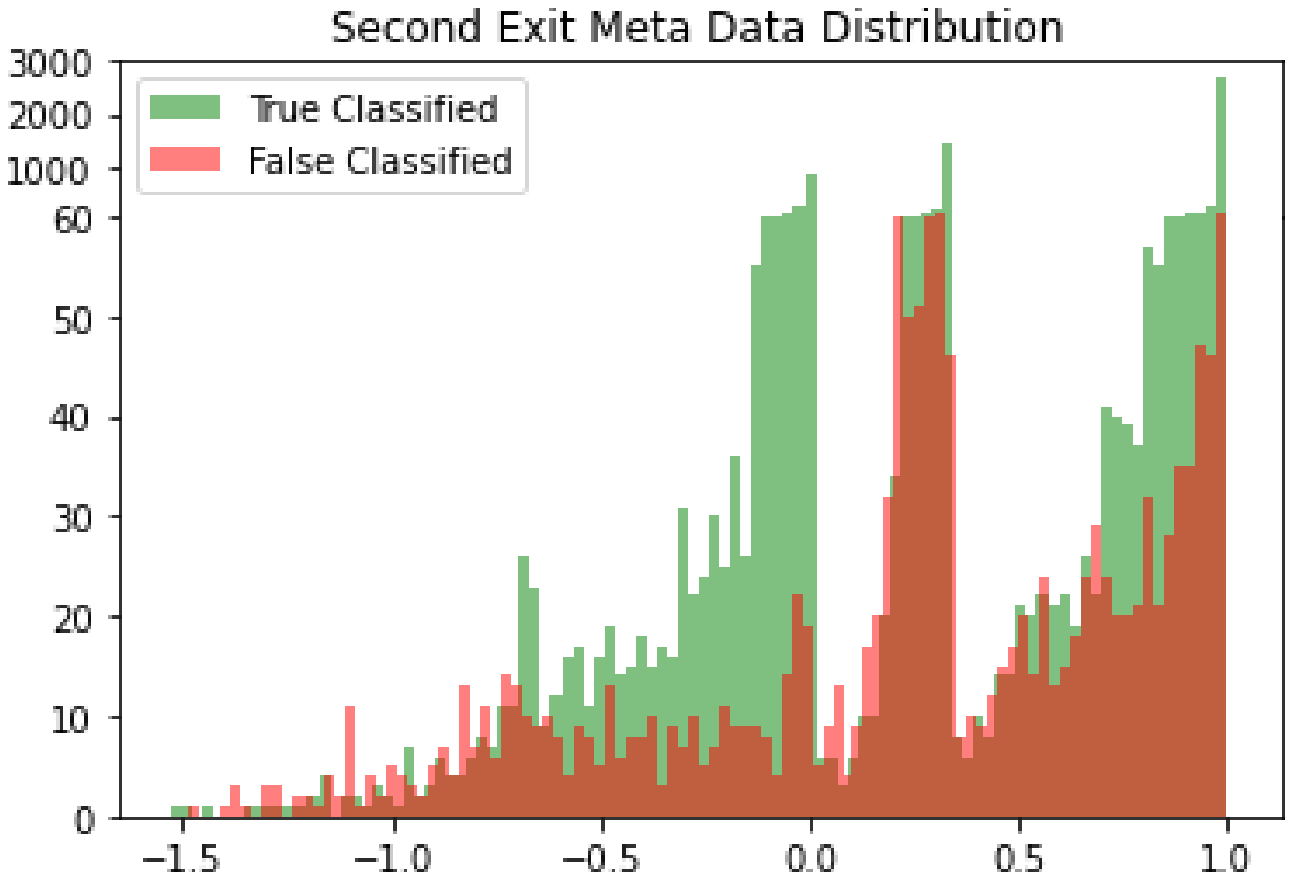}
    \caption{Second Exit Meta Data Distribution}
  \end{subfigure}
  
  \caption{Distribution of different meta information extraction from validation set after being classified }
  \label{fig:metaInformation}
\end{figure*}

In the Split-AI framework, the decision unit is responsible for identifying samples with high uncertainty by extracting meta-information from the classification results. It can be a machine learning model that has learned true classified and false classified samples based on meta-information. In fact, extracted meta information from each sample can be considered as features and whether the sample has been classified truly can be used as a label to train such a binary classifier. We extracted meta information from validation set data and used them to train such a classifier and evaluate a decision unit using the test set data. Figure~\ref{fig:metaInformation} shows the distribution of meta-information after applying all the validation set data to the client model. It can be easily seen that meta-information has a good separability feature based on true and false classified samples. For a machine learning classifier, two different models have been trained. One model is the deep neural network model with 4 hidden layers and another is an XGBoost model\cite{xgboost} using a gradient boosting algorithm. Our experimental result shows that the deep learning model performs a little bit better on the test set.

\begin{table}

\centering
\caption{Example of meta information. probability of each sample for different classes and their related meta information using equation \ref{second_eq}-\ref{fifth_eq}}
\label{table:metainformation}
\begin{tabular}{|c|c|c|c|}

\hline
      &\textbf{sampleA} & \textbf{sampleB} \\ \hline
      \textbf{class1}& 0.9 & 0.2 \\ \hline
      \textbf{class2}& 0.09&0.5 \\ \hline
      \textbf{class3}&0.01&0.3 \\ \hline
      \textbf{probability} & 0.9 & 0.5 \\ \hline
      \textbf{least confidence} &  0.81& 0.2\\ \hline
      \textbf{entropy} & -0.35 & -1.02\\ \hline
      \textbf{standard deviation} & 0.12 & 0.4 \\ \hline
      
\end{tabular}
\end{table}

\subsection{The Client Model}\label{sec:Client Model} 
Designing a neural network for IoT devices has absorbed quite a little attention during recent years. Researchers are trying to come up with new models that have a low storage and inference time and higher accuracy. The research on this topic can be divided into two main methodologies. In the first method, a light model with a good accuracy has been designed to solve the classification, segmentation, and object detection problem.  SqueezNet\cite{squeezenet}, MobileNet\cite{mobilenets}, MobileNetV2\cite{mobilenetv2}, EfficientNetB0 and B1\cite{EfficientNet}, MNasNet\cite{mnasnet} are the architectures that have been introduced recently. In the second method, Researchers are trying to reduce the model size of big architectures while maintaining their accuracy. Techniques like compression\cite{compressionsurvey} and distillation\cite{kd-survey} are being used in recent years to take advantage of famous architectures. 

In our proposed algorithm, we have harnessed both methods simultaneously. In this manner, unlike the other compact DNN models which an expert has designed a light network for IoT devices, the model can be obtained automatically. Moreover, In the knowledge distillation technique, the student model needs to follow the structure of the cumbersome model to get the best performance, but in the proposed method, a simpler model does not need to follow any specific structure and reaching the good accuracy with acceptable model size is the final goal. Moreover, since the performance of the decision unit is highly dependent on the extracted meta information from the client model, it is really important to have a client model with an acceptable accuracy.

To apply the proposed method, first, the NAS morphism algorithm has been applied to the train data set to find the architecture and train it using the softened labels. In the morphism algorithm, the algorithm is using validation loss to evaluate each step. In the proposed method, during the searching process, instead of training the model with hard labels, softened labels from a cumbersome model have been used. The softened loss also has been used to evaluate each iteration. Since hill climbing is taking a random step in each iteration to find the maximum point in the search space, by stronger evaluation of each random step, the algorithm is more likely able to find the better way to reach the top of the hill. In this manner, by transferring the information from a cumbersome model and evaluate each search iteration based on the softened loss, the algorithm is more able to find the better architectures and layers to reach the cumbersome model's accuracy. To avoid getting a big model, the iteration of the hill-climbing greedy algorithm is set to a small value.

For producing the soft labels, using the exact method in \cite{hadi}, some imageNet\cite{imagenet} pre-trained state-of-the-art neural network architecture such as InceptionResNet\cite{inceptionv4}, Xception\cite{Xception}, and DensNet\cite{DenseNet} have been trained by using the transfer learning method\cite{transfersurvey}. Among them, the InceptionResNet model has been used to produce the softened label since it outperforms others by the accuracy of $86.5\%$. The temperature value($T$) first is set to $5$, but the accuracy and model size did not change a lot comparing to the hard label search. So, we considered $T=20$ for obtaining more softened labels. After finding the architecture, hard labels are used to train the model from scratch. The obtained model is converted to a two-exit model by adding an exit layer to the middle of the model's architecture.

\subsection{Multi-exit architecture}\label{sec:Multi-exit}
The proposed framework can be extended to three stages. The client model can be converted to a two-exit neural network and a similar idea for the last exit can be applied to it. The output logits of earlier exit have been calculated after applying the model to the validation set data set and after extracting the meta information, a separate deep neural network model has been trained as a decision unit for this stage. Figure \ref{fig:metaInformation} (a) shows the meta information extracted from the first exit.
Based on meta information, this time, true classified samples from the earlier exit can be determined by the decision unit and show the classification result to the users. By doing so, a sample does not need to be calculated through all the layers of the neural network in the client model on IoT devices which is computationally efficient. Consequently, the inference time can be reduced and more energy can be saved.
\section{Experimental Result}\label{sec:result}
Machine learning models are very appropriate to recognize hard diseases such as breast cancer recognition\cite{breast_cancer}, skin lesion\cite{skin_solo}.
These models are providing the opportunity for the majority of people to observe their well-being in an easy, accessible manner. Consequently, a variety of healthcare frameworks have been implemented on mobile devices \cite{farahanihealthcare}. Consequently, we have used the international Skin Imaging Collaboration Challenge 2019
(ISIC 2019) \cite{skindata_1, ISIC_2, skin_data_2} to evaluate our proposed framework. The total number of images contain train and test images is 33,569 images. nevertheless, only the labels of the training data are available. In this paper, only the labeled images have been harnessed to evaluate the proposed methods. The number of labeled images is 25,331 images of eight different skin diseases which are basal cell carcinoma, benign keratosis, vascular lesion, melanoma, squamous cell carcinoma, melanocytic nevus, actinic keratosis, and dermatofibroma. The number of images for each classes are 3,323, 2,624, 253, 4,522, 628, 12,875, 876, 239 respectively. We randomly split 80\%, 10\%, and 10\% as training data, validation data, and testing data respectively.

\subsection{Experimental Considerations}
For experimental evaluation, Python has been used. For the deep learning approach, Keras which is a platform that is running on top of the TensorFlow has been used\cite{keras_doc}. All the obtained models can be converted to the TensorFlow lite format to implement on real devices. A single GPU (Nvidia GeForce GTX 1080 Ti with 11 GB GDDR5X memory) has been used for all the evaluations.

\subsection{Evaluating the performance of KD-NAS}
Results of applying the knowledge distillation technique to the search algorithm of the NAS morphism show that the proposed method is able to find an appropriate model for implementing on mobile devices compared to the state-of-the-art light models. Table ~\ref{table:kd-nas} shows the experimental results. The NAS model is the model that has been obtained by applying the method in \cite{nas-morphism} to the data set. The KD-NAS model has been obtained by applying the knowledge distillation technique during the search time and using the softened loss to evaluate each iteration.

The iteration number has been set to a small number to avoid expanding and obtaining a big model. After finishing the searching procedure and finding the model, the model was trained from scratch. For training other models on the table, the imageNet\cite{imagenet} pre-trained architectures have been used and models have been trained using the transfer learning method. The InceptionResNet\cite{inceptionv4} model is a complex model with higher accuracy and it has been used as a server model in this paper. All other models are state-of-the-art models that have been proposed by researchers for deploying on mobile and IoT devices. Among them EfficientNet-B0 and EfficientNet-B1\cite{EfficientNet} have been obtained using the neural architecture search technique and MobileNet\cite{mobilenets} and MobileNetV2 \cite{mobilenetv2} are human-designed models. Experimental results show that even though the NAS and Specially KD-NAS are trained from scratch, they have good accuracy and an acceptable number of parameters to be deployed on the IoT devices.

\begin{table}

\centering
\caption{KD-NAS and other state-of-the-art architectures.}
\begin{tabular}{|c|c|c|c|}

\hline
      \textbf{Architecture} is &\textbf{Accuracy}& \textbf{Number of Parameter} \\ \hline
      \textbf{EfficientNet-B0\cite{EfficientNet}}& 81.5 & 4,059,812 \\ \hline
      \textbf{EfficientNet-B1\cite{EfficientNet}}& 83&6,585,480 \\ \hline
      \textbf{MobileNet\cite{mobilenets}}&80.5& 3,237,064 \\ \hline
      \textbf{MobileNetV2 \cite{mobilenetv2}} & 76 &  2,268,232 \\ \hline
      \textbf{InceptionResNet \cite{inceptionv4}} &  85.8 & 54,877,872\\ \hline
      \textbf{NAS\cite{nas-morphism}} & 80 & 1,090,312\\ \hline
      \textbf{KD-NAS} & 82 & 4,969,032 \\ \hline
      
\end{tabular}
\label{table:kd-nas}
\newline
\end{table}

\begin{table}

\centering
\caption{KD-NAS and other state-of-the-art architectures decision unit performanc.}
\begin{tabular}{|c|c|c|c|c|}

\hline
      \textbf{Architecture}  &\textbf{Accuracy}& \textbf{AUC} & \textbf{Sensitivity} & \textbf{Specificity}\\ \hline
      \textbf{EfficientNet-B0\cite{EfficientNet}}& 81 & 0.79 & 0.91 & 0.32\\ \hline
      \textbf{EfficientNet-B1\cite{EfficientNet}}& 82 & 0.78 & 0.94& 0.19 \\ \hline
      \textbf{MobileNet \cite{mobilenets}}&83& 0.80 & 0.95 & 0.26\\ \hline
      \textbf{MobileNetV2 \cite{mobilenetv2}} & 75 &  0.74 & 0.8&0.51 \\ \hline
      \textbf{KD-NAS} & 84 & 0.83 & 0.95&0.32 \\ \hline
      
\end{tabular}
\label{table:kd-nas-du}
\newline
\end{table}

\subsection{Evaluate the performance of Split-AI (Decision Unit)}
\begin{figure}
  \centering
    \includegraphics[width=\linewidth]{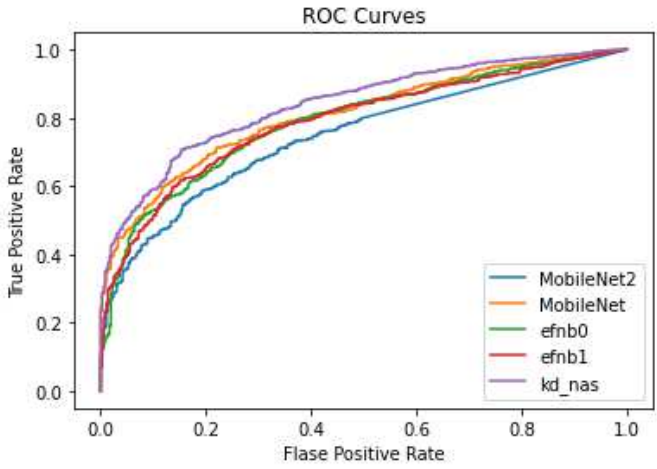}
     
  \caption{The ROC Curves of Decision Units}
  \label{fig:ROC_du}
\end{figure}

For evaluating the performance of our proposed Split-AI framework, we have focused on the performance of the decision unit. To do so, after training the model using the train data set(the result of the table ~\ref{table:kd-nas}), the output logits of the client model have been calculated by applying the model to the validation data set and meta information have been extracted from logits as features for training the decision unit model. The decision unit is a binary classifier model which classifies the test sample based on whether sending it to the server or not. Four trained models from table~\ref{table:kd-nas} (EfficientNet-B0, EfficientNet-B0, MobileNet, MobileNetV2), as well as the KD-NAS model (proposed model), have been used to extract meta information, and a separate decision unit classifier has been trained using each model's meta information for comparison. Figure ~\ref{fig:ROC_du} shows the ROC curves of the different decision unit models for different client models on table~\ref{table:kd-nas}. Experimental results show that the decision unit classifier of the proposed client model outperforms the decision unit classifier of the state-of-the-art models in terms of accuracy and AUC(area under the curve) number. Table ~\ref{table:kd-nas-du} shows the performance of the decision unit classifiers of different client models. 

\begin{figure}
  \centering
    \includegraphics[width=\linewidth]{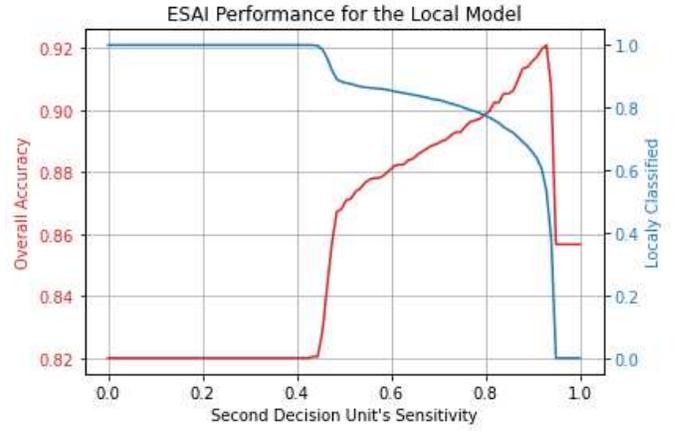}
     
  \caption{The performance of ESAI system with only second decision unit, x-axis is the sensitivity of the second decision unit}
  \label{fig:performance_saia}
\end{figure}

For evaluating how the decision unit idea can improve the overall accuracy, we also investigated the influence of the decision unit on the overall accuracy in Figure \ref{fig:performance_saia} by evaluating only the second(normal) exit's performance based on different decision unit's sensitivity numbers. As it can be seen, when the sensitivity is the lowest ($0$), it means that all the samples are being classified by the local model and the accuracy is around $82\%$ and when the sensitivity is highest($1$), all samples are being classified by the server model and the accuracy is around $86\%$ . For the sensitivity between $0.5$ and $0.9$, the overall accuracy is better than the client and server models and $92\%$ accuracy can be reached by only sending around $40\%$ of the test samples to the server which shows the effectiveness of the decision unit. For calculating the overall accuracy this equation has been used:
\begin{equation}
acc = \frac{T_{N} + S_P}{S_T}   
\end{equation}
Where $T_{N}$ is the true negative number of the second exit, $S_P$ is the number of samples that have been sent to the server and have been classified correctly and $S_T$ is the total number of test samples. For the server model, the image-net pre-trained inceptionResnet\cite{inceptionv4} model in table~\ref{table:kd-nas} has been used. Any models with better performances can be replaced by this model to get better accuracy in the infrastructure implementation, notwithstanding.

\subsection{Evaluating the Performance of  Multi-exit Architecture}
To implement the multi-exit architecture, the output probability of the first exit decision unit has been calculated. If the certainty of the classification task is higher than the decision unit's sensitivity, the sample will be classified at the current exit, otherwise, it will be sent to the next exit. The same procedure will be applied to the last exit. 
For those samples that have been chosen to be sent to the server, the classification result of the server model has been calculated for each of them. The total accuracy has been calculated based on the following equation:
\begin{equation}
acc = \frac{T_{N1} + T_{N2} + S_P}{S_T}   
\end{equation}

\begin{figure}
  \centering
    \includegraphics[width=\linewidth]{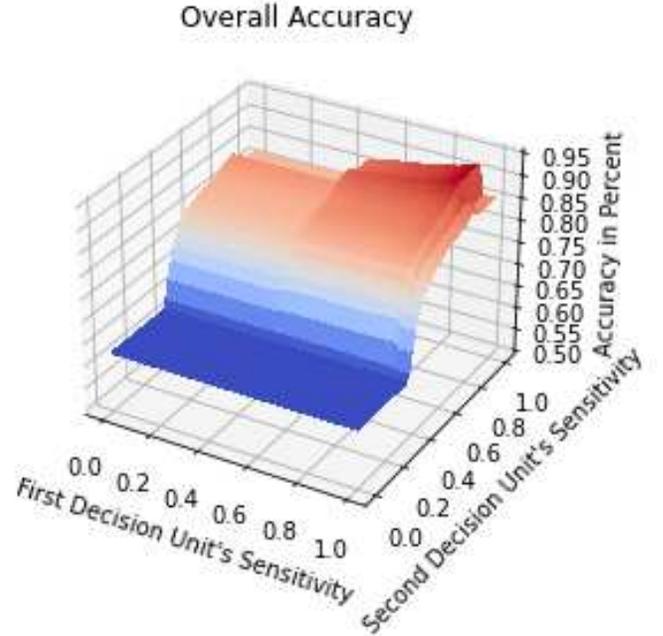}
     
  \caption{The accuracy of ESAI system for different sensitivities, x-axis is the values of sensitivity for the first decision unit and y-axis is the values of sensitivity for the second decision unit.}
  \label{fig:accuracy_saia}
\end{figure}

where $T_{N1}$ is the true negative number of the first exit, $T_{N2}$ is the true negative number of the second exit, $S_P$ is the number of samples that have been sent to the server, and have been classified correctly and $S_T$ is the total number of test samples. In this way, the whole accuracy of the framework has been calculated regardless of in which stage the samples are going to be classified. 

\begin{figure*}[h!]
  \centering
  \begin{subfigure}[b]{0.4\linewidth}
    \includegraphics[width=\linewidth]{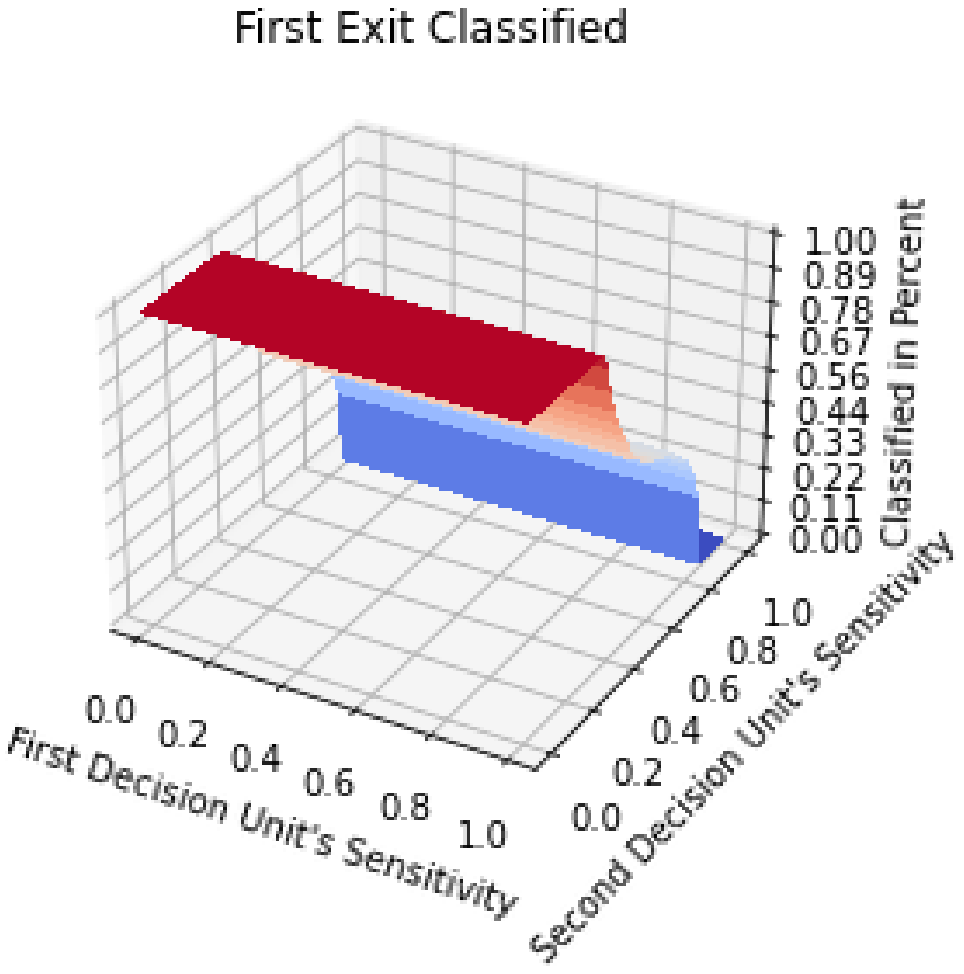}
     \caption{First exit classified samples}
  \end{subfigure}
  \begin{subfigure}[b]{0.4\linewidth}
    \includegraphics[width=\linewidth]{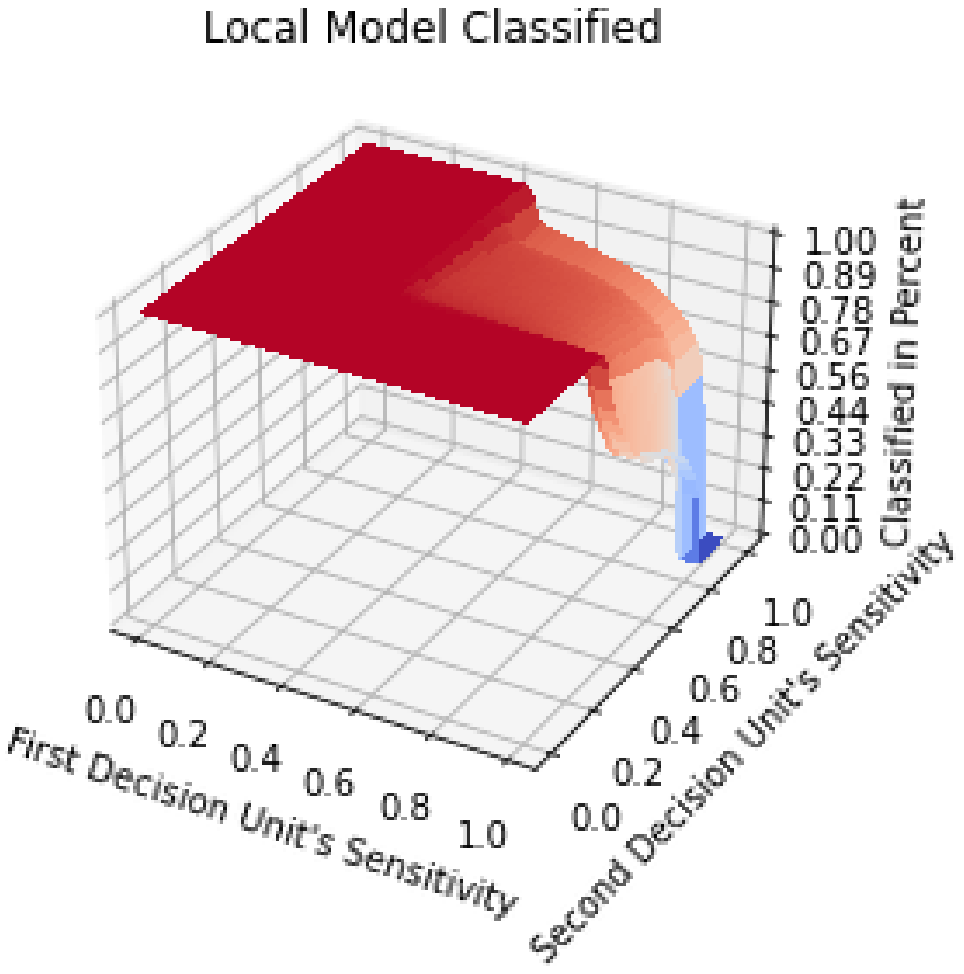}
    \caption{Local model classified samples}
  \end{subfigure}
  
  \caption{Number of classified samples from the first exit and the on-device model based on different values of sensitivity  }
  \label{fig:classified_samples}
\end{figure*}

Figure~\ref{fig:accuracy_saia} and Figure \ref{fig:classified_samples} show the accuracy of the whole Split-AI framework and the number of samples that have been classified on the local model and its earlier exit based on a different amount of sensitivities of each decision unit.  According to Figure~\ref{fig:accuracy_saia}, the model accuracy is changing from $63.5\%$ to $92\%$.  If all the samples have been classified in the first exit, the accuracy is $63.5\%$. If they have been classified using only the second exit, the accuracy is $82\%$, and using the server model for all samples, the accuracy would be $85\%$. When the accuracy of the model is around $90\%$, $60\%$ of the test data has been classified in the client model and only $40\%$ of them have been sent to the server. From the samples that have been classified on the client, $18\%$ of them have been classified in the first exit. This demonstrates the efficacy of our proposed method which can bring more communication (sending fewer data to the server) and computation (less inference time by exiting from the first exit) energy saving for the mobile and IoT devices, while the overall accuracy has been improved.    

\subsection{Evaluating the effectiveness and efficiency of our proposed system}
We implemented the whole framework on real devices. For doing so, after training the networks, we have three networks from the beginning to the dividing node (Figure \ref{fig:node_one}), from the node to the first exit (Figure \ref{fig:exit_one}), and from the node to the last exit (Figure \ref{fig:exit_last}). We also have two decision unit networks. After converting all the five networks to the TensorFlow lite, the whole framework has been implemented on real devices. Samsung-s10 devise and iPhone-11have been used to implement the framework on them as an 
Android and IOS platforms respectively. Table \ref{table:multi_exit_mobile} shows the inference time of the framework on each mobile platform. According to the table, the inference time of the second exit is almost twice the inference time of the first exit for both mobile platforms. It is also possible to tune the sensitivity of each decision unit to regulate the amount of data that should be sent to the last exit and to the server based on device battery status and communication network availability. The framework can use both the client and the server model in normal conditions. If the battery power is not really high or the satellite is impeded, the framework can use only the client model, and if the battery power is really low, the framework can just use the first exit for the classification.

\begin{table}

\centering
\caption{Inference time of the ESAI framework on real devices in Millie Second.}
\begin{tabular}{|c|c|c|c|c|}

\hline
      \textbf{Device}  &\textbf{First Exit}& \textbf{Second Exit} \\ \hline
      \textbf{Samsung-s10(Android)}& 38 & 74\\ \hline
      \textbf{Iphone-11(Ios)}& 34 & 68 \\ \hline

\end{tabular}
\label{table:multi_exit_mobile}
\newline
\end{table}

\section{Related Work}\label{sec:related}

\subsection{Compact Deep Neural Networks} \label{sec:relatedWork_DL_Mobile}
Compact Deep Neural Networks (Compact DNNs) approaches have been proposed to enable the development and deployment of real-world AI applications (e.g., mobile healthcare, smart home, wearable technologies) on mobile and IoT devices. Most of such compact DNNs approaches build efficient DNNs by creating efficient building blocks for removing the redundancy of the current DNNs designs. 

For instance, by downsampling the data using $1\times1$ convolution filters, SqueezeNet \cite{squeezenet} obtains AlexNet \cite{alexNet} level of accuracy with 50x fewer parameters and less than 0.5MB model size. 
MobileNet \cite{mobilenets, mobilenetv2, howard2019searching} significantly reduces computation complexity without accuracy loss, compared with traditional DNN models, by suggesting a useful building block, ``inverted residual block'' into its design of DNNs. 
By using customized architecture, that only has one forth operations of VGG16 \cite{simonyan2014very}, YOLO \cite{redmon2016you} becomes one of the state-of-the-art, real-time object detection systems. 
Recently, efficientNet \cite{EfficientNet} has been proposed for execution on mobile and IoT devices, that uniformly scales each dimension (e.g., width, depth, and resolution) of DNN models with a fixed set of scaling coefficients. 

Although the compact DNNs could dramatically reduce the computation complexity, such ``hand-craft'' approaches require substantial design insight to optimized the compact DNNs' performance. And the overall performance of the compact DNN model still would not be as good as the more advanced models deployed on the server-side, which could also be the ensemble/fusion of several well-trained DNN models. 

\subsection{Compressed Deep Neural Networks} \label{sec:relatedWork_Compressed_DNN} 
DNN model compression \cite{hinton-distilling, ba2014deep, polino2018model, huynh2017deepmon, han2016eie, han2015deep, liu2018demand, zhao2018deepthings} is an alternative, automatic approach that does not require certain prior design principles to design efficient DNNs running on mobile and IoT devices. For instance, data quantization \cite{han2015deep} has been proposed to reduce the number of bits to represent each weight value of DNN models. 
Network pruning \cite{luo2017thinet} has been proposed to trim the network connections within DNNs that have less influence on the inference accuracy. 
Knowledge distillation \cite{hinton-distilling, ba2014deep, polino2018model} has been proposed to compress a model by teaching a simplified student DNN model, step by step, exactly what to do using a complex pre-trained teacher DNN model and then deploy the student DNN model on the mobile devices. 
Compressed DNNs provide more flexibility in designing efficient DNNs running on mobile and IoT devices. However, solely deploying compressed DNN models on mobile and IoT devices cannot take advantage of the more advanced models deployed on the server-side.

\subsection{Split Deep Neural Networks} \label{sec:relatedWork_Split_DNN} 
Split-DNN architectures \cite{jeong2018computation, kang2017neurosurgeon, lane2016deepx, osia2020hybrid} have been proposed to split a DNN into the client-side component and the server-side component, that maintains the lightweight feature extraction and data prepossessing DNN functionalities on the mobile or IoT devices, and pushes the execution of complex DNN models to much more powerful servers. 
For instance, Osia et al. \cite{osia2020hybrid} proposes a hybrid architecture where a DNN model, that has previously been trained and fine-tuned on the cloud, would be split into two smaller neural networks: a feature extraction network that runs on the mobile or IoT devices, and a classification network that runs on the cloud system, and both neural networks on the local device and the cloud system would collaborate on running the original complex DNN model. 
Matsubara et al. \cite{matsubara2019distilled} propose a KD-based Split-DNN framework to reduce the communication cost between the client and the server. However, such approaches usually cannot fully rely on the client-side model, thus unable to work if the communication is impeded. 
However, none of such approaches consider the communication bottleneck between the client and the server. Also, the existing approaches cannot adjust the split-AI work assignments between the client and the serve depending on the device's condition (e.g., storage size,  power consumption, and communication bandwidth).

\subsection{Multi-exit Deep Neural Networks} \label{sec:relatedWork_MultiExit_DNN} 
Multi-exit Deep Neural Networks \cite{teerapittayanon2016branchynet, huang2017multi, kim2018doubly, zhang2018graph, phuong2019distillation} are proposed to design DNNs with additional exits, where an appropriate exit would be chosen at the testing phase in terms of certain criteria (e.g., accuracy, efficiency, and power consumption). 
For instance, 
BranchyNet \cite{teerapittayanon2016branchynet} designs their early exiting structures of DNNS by augmenting existing standard DNN architectures such as LeNet, AlexNet, and ResNet. 
Multi-Scale \cite{huang2017multi} DenseNet is a customizable multi-exit architecture inspired by the idea that the network structure of earlier exits could serve as the feature extractors for the later exits. 
DNNet \cite{kim2018doubly} is a doubly nested network where each neuron represents a single sub-model that all aim to solve the same task (i.e., image classification). 
Compared with the existing multi-exit DNN approaches, our approach introduces network morphism-based Neural Architecture Search (NAS) together with Knowledge Distillation techniques into the design of multi-exit split-DNN.

{\bf Our work:} Our proposed split-AI architecture takes advantage of not only the compact and efficient DNNs on the client-side but also the highly accurate DNNs on the server-side, thus, outperforms just applying the compact DNNs. For instance, we designed a novel split-AI architecture to enable efficient and accurate deep learning services from both the mobile/IoT devices and the server. We introduced Knowledge Distillation techniques into the searching strategy of morphism-based NAS to produce ``lighter'' DNN models. And we proposed to utilize multi-exit DNN architecture to enable flexible and efficient trade-off tuning between the resource usage efficiency and the prediction accuracy.

\section{Conclusion}
In this paper, a novel and effective framework for the mobile healthcare system has been proposed. This framework is able to get benefits of both models on the cloud server and the client model intelligently.
To do so, a framework contains a decision unit model that is responsible to make a decision about each sample to whether sending it to the server or not. After classifying the model by the client model, the decision unit tries to intelligently evaluate the exactitude of the classification task and send the samples with high classification uncertainty. Early exiting classification has been produced by following a similar idea in which a decision unit makes the decision on the classified samples with a lower uncertainty in the earlier exit to reduce the computational power. To obtain an accurate and acceptable model, the knowledge distillation technique has been applied to the morphism-based neural architecture search methodology in a specific way and then define an early exit before the last convolutional layers of the model. In our future work, we will try to enhance the framework by introducing better decision unit techniques and more powerful meta information. Also, we will try to Improve the model parameters based on the last decision on the sample in an intelligent manner.


\section*{Acknowledgments} Effort sponsored in part by United States Special Operations Command (USSOCOM), under Partnership Intermediary Agreement No. H92222-15-3-0001-01. The U.S. Government is authorized to reproduce and distribute reprints for Government purposes notwithstanding any copyright notation thereon. \footnote{The views and conclusions contained herein are those of the authors and should not be interpreted as necessarily representing the official policies or endorsements, either expressed or implied, of the United States Special Operations Command.}

\bibliographystyle{IEEEtran}
\bibliography{IEEEabrv,main_text}

\section{Appendix}
\begin{figure*}[h!]
  \centering
    \includegraphics[width=0.70\linewidth]{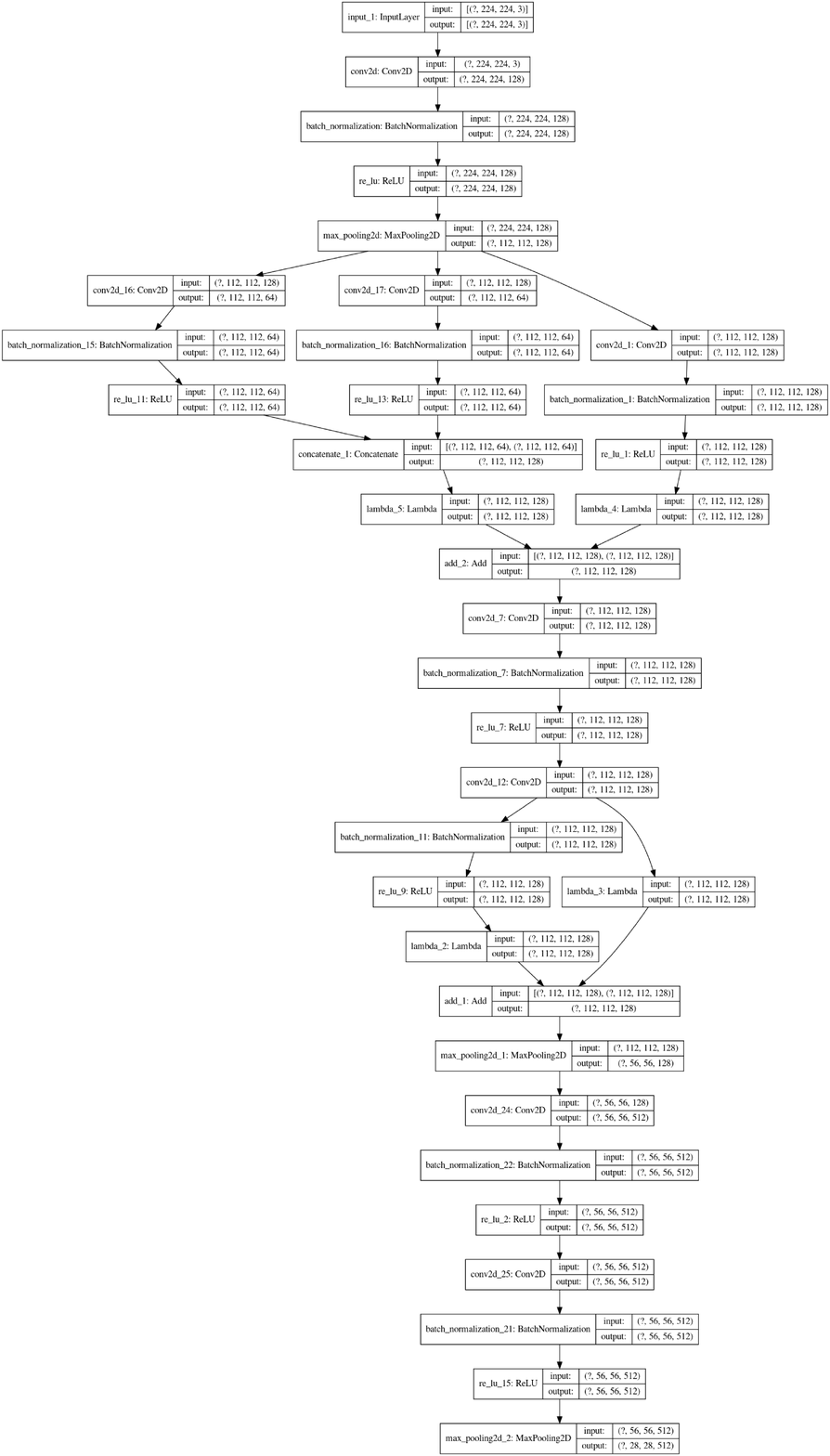}
  \caption{Initial section of the client model obtained using proposed NAS algorithm.   }
  \label{fig:node_one}
\end{figure*}

\begin{figure*}[h!]
  \centering
    \includegraphics[width=0.75\linewidth]{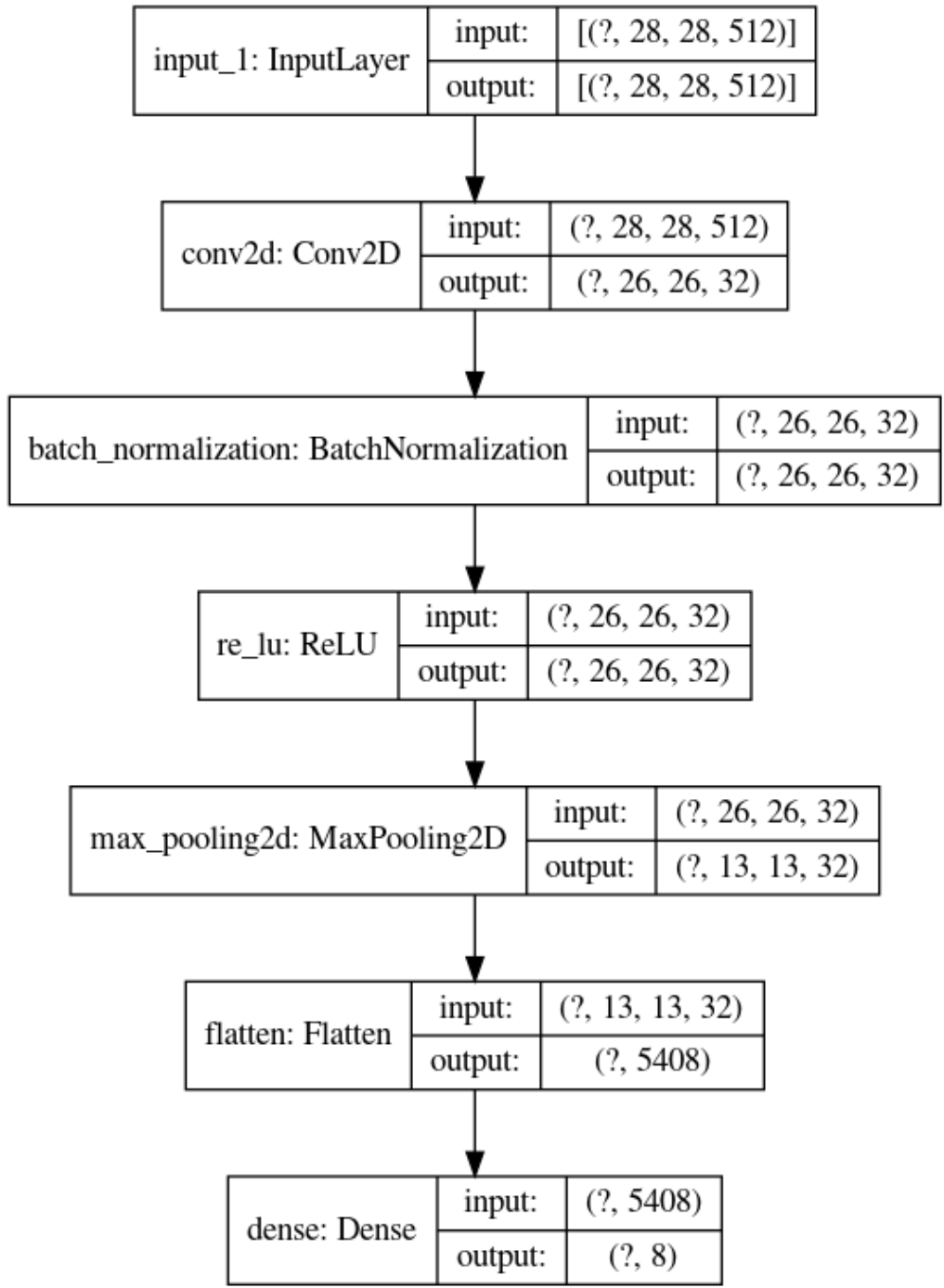}
  \caption{First Exit of the client model with manual design.   }
  \label{fig:exit_one}
\end{figure*}

\begin{figure*}[h!]
  \centering
    \includegraphics[width=0.75\linewidth]{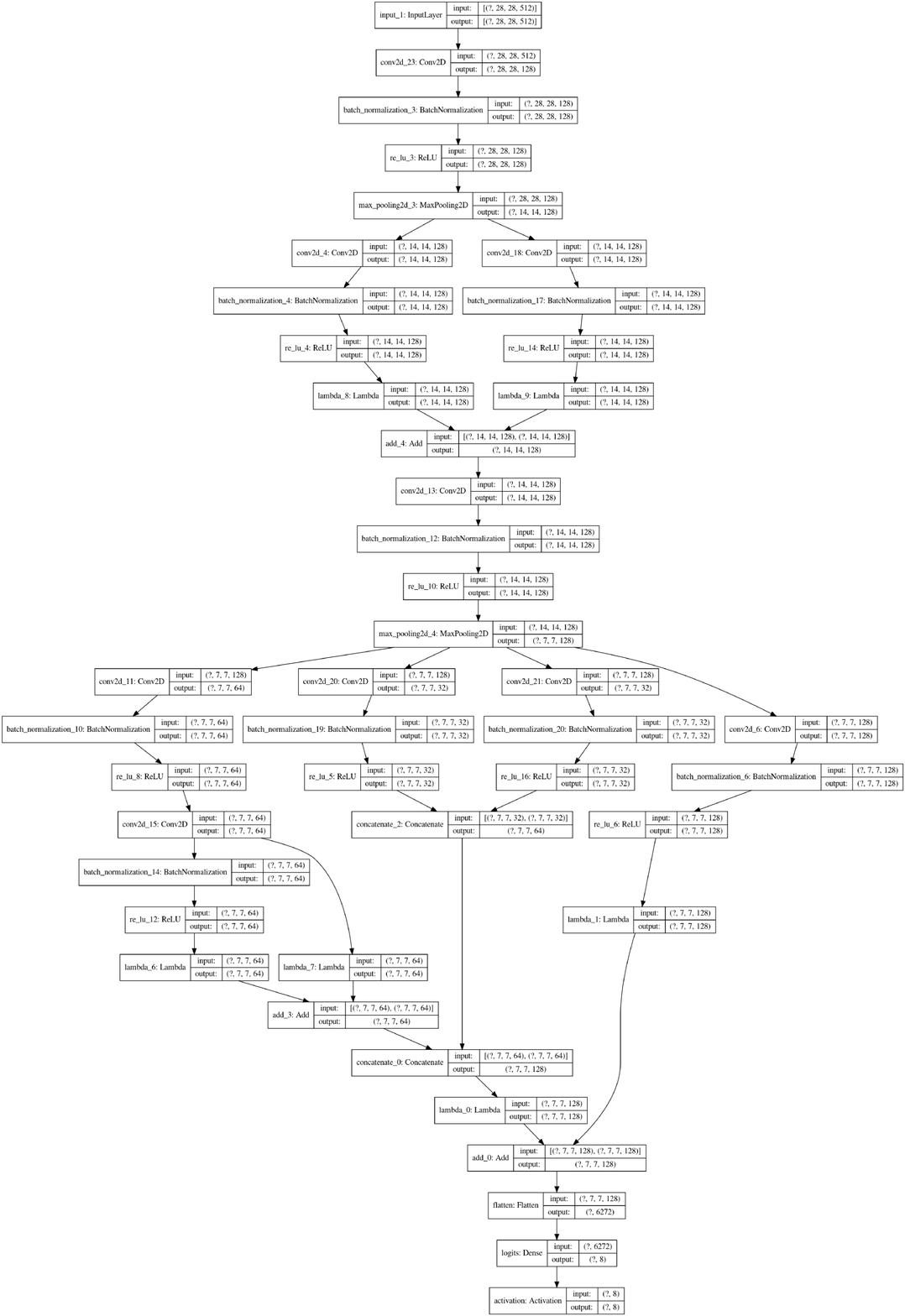}
  \caption{Second Exit of the client model obtained using proposed NAS algorithm.   }
  \label{fig:exit_last}
\end{figure*}

In this section, the client network has been depicted. After applying the proposed NAS method to the data set. An earlier exit has been applied manually to the network. Figure \ref{fig:node_one} shows the initial section of the network from input to the divided node. Figure \ref{fig:exit_one} shows the architecture from dividing node to the earlier exit. Figure \ref{fig:exit_last} shows the rest of the network from the divided node to the normal output.

\end{document}